# Tachyons Before Tachyons: Lev Strum (1890-1936) and Superluminal Velocities

Helge Kragh[*]

**Abstract:** No particle or signal carrying information can travel at a speed exceeding that of light in vacuum. Although this has for a long time been accepted as a law of nature, prior to Einstein's 1905 theory of special relativity the possibility of superluminal motion of electrons was widely discussed by Arnold Sommerfeld and other physicists. Besides, it is not obvious that special relativity rules out such motion under all circumstances. From approximately 1965 to 1985 the hypothesis of tachyons moving faster than light was seriously entertained by a minority of physicists. This paper reviews the early history concerning faster-than-light signals and pays particular attention to the ideas proposed in the 1920s by the little-known Ukrainian physicist Lev Strum (Shtrum). As he pointed out in a paper of 1923, within the framework of relativity it is possible for a signal to move superluminally without violating the law of causality. Part of this article is devoted to the personal and scientific biography of the undeservedly neglected Strum, whose career was heavily – and eventually fatally – influenced by the political situation in Stalin's Russia. Remarkably, to the limited extent that Strum is known today, it is as a literary figure in a novel and not as a real person.

## 1. Introduction

The velocity of light in vacuum is a magical quantity. It has puzzled physicists and philosophers ever since Einstein in 1905 postulated that this particular velocity is a natural constant independent of the state of motion of the light source. The velocity of light had a special status even in pre-relativity electromagnetic physics as it was generally agreed that a material body cannot move with a velocity equal to or greater than $c$. Nor can a communicative signal be transmitted superluminally, with a velocity exceeding that of light or, generally, electromagnetic waves whatever their wavelength. (In this paper, 'velocity of light' refers throughout to motion in *vacuum* and not in some optical medium.) The possibility of signal transmission at $v > c$, or the motion of electrons at such velocity, was discussed until about 1905, but with the acceptance of special relativity the discussions largely ceased. According to a monograph on superluminal motion, "the whole problem lost its appeal, and was abandoned for about half a century" [1, p. 166]. But, as we shall see, this is to oversimplify history.

Are superluminal velocities really ruled out by the theory of relativity such as claimed by Einstein and other experts in this branch of fundamental physics? A few physicists, none of them well known either then or today, questioned the consensus view. Without suggesting that superluminal signals are real phenomena actually found in nature, they argued that they are consistent with Einstein's theory and in this sense are possible phenomena. This is what the Russian-Ukrainian physicist Lev Strum concluded in a series of papers from the 1920s but without his work making any impact on the physics community. Nor has it been noticed by historians of modern physics[1] who may have been unaware of a pioneering paper written on the subject by the Russian physicist Grigorii Malykin and two co-authors in *Physics-Uspekhi* [3].

This paper examines the early faster-than-light story with a focus on the works of the little-known Strum and his tragic fate in Stalinist Russia where he was executed in 1936. When the study of hypothetical particles travelling faster than light was revived in the 1960s, soon to be called tachyon physics, none of the involved physicists were aware of Strum, a physicist from the past who was not only obscure but completely forgotten. Although his works on superluminal signals were published in mainstream physics journals and anticipated important parts of the later tachyon theory, they failed to attract attention. Following an introductory sketch of the modern hypothesis of tachyons ca. 1960-1980, the paper outlines the somewhat similar hypothesis as discussed in pre-relativistic electron physics (Section 2) and within the framework of Einstein's theory of special relativity (Section 3). It ends with three sections devoted to Strum, two on his life and career – and literary resuscitation – and one on his works dealing with superluminal signals.

## 2. Tachyons

A *tachyon* is a hypothetical particle (or signal) that moves superluminally, that is, at a speed greater than that of light in empty space, $v > c$. The name, a neologism, was coined by American physicist Gerald Feinberg in 1967 from the Greek word *takhys* (ταχύς) meaning swift or fast [4]. The synonymous 'faster-than-light particle' had previously been employed and can still be found, sometimes with faster-than-light abbreviated to FTL. Another synonym employed by Arthur Miller [5] and a few other writers is the adjective 'hyperlight.' While Feinberg did not suggest a new word for normal particles moving at $v < c$, two years later *tardyon* entered the physicists' vocabulary [6], derived from the Latin 'tardus' meaning slow. Both terms are accepted by the *Oxford English Dictionary* [7].

---

[1] In a book of 2011 I referred briefly to "an unknown Russian physicist by the name of L. Strum … [whose superluminal velocity] argument went unnoticed and only reappeared in the 1960s" [2, p. 298].

Interestingly, at least from a philological point of view, the now commonly used term 'superluminal' only turned up in an English-language physics text in 1969 followed by 'subluminal' two years later [7]. Hyphenated as 'super-luminal' the adjective first appeared in the philosopher Karl Popper's influential *The Logic of Scientific Discovery* published ten years earlier [8, p. 236], [9, p. 112 and p. 144]. Popper coined the term as an English-Latin translation of the word *Überlichtgeschwindigkeit* as it appeared in the German original *Logik der Forschung* from 1935. This rather cumbersome word was routinely used (and is still used) in the German physics literature where it since the 1890s appeared in discussions of signals propagating at speeds exceeding that of light (Section 3).

The consensus view until the 1960s was that superluminal particles were inconsistent with Einstein's theory of relativity and therefore belonged to either philosophy or to science fiction. It was a popular topic of after-dinner conversation, but not one of scientific interest. However, in 1962 three American physicists at the University of Rochester argued convincingly that particles moving faster than light are in fact compatible with special relativity theory [10]. They suggested that the usual objections against the idea were invalid if only the 'meta particles' (as they called them) were assumed to be created with $v > c$ and remained in this state forever. According to the three physicists, the hypothetical meta particles could not be directly observed but indirectly they might be detected and thus turned into real particles. The 1962 paper attracted only modest attention until Feinberg developed the idea into an elaborate quantum field theory of what he called tachyons. According to Feinberg [4]:

> The possibility of particles whose four-momenta are always spacelike, and whose velocities are therefore always greater than c is not in contradiction with special relativity, and such particles might be created in pairs without any necessity of accelerating ordinary particles through the 'light barrier.'

The study of faster-than-light particles caught on and peaked in 1978 when 44 scientific papers were published with 'tachyon' or some variant thereof in the title.

Several of these papers reported experimental searches, but alas, all of them with negative results.[2]

Physicists involved in tachyon theory agreed that the particles, if they existed, had most peculiar properties. They have a mass and are possibly electrically charged, but the mass is imaginary ($m^2 < 0$) and therefore not measurable. A directly measured physical quantity can only be expressed by a real number [1, pp. 226-235]. As a photon in empty space cannot be brought to rest so a tachyon will always live in the $v > c$ regime. It can be slowed down, but never to the velocity of light or below it. Strangely, the speed of a tachyon increases as its energy decreases, just the opposite of what is the case for ordinary particles. To slow down a tachyon to $v = c$ will require an infinite amount of energy just as it takes infinite energy to boost a normal subluminal particle to this limit.

While such behaviour is strange indeed, it is even stranger that in a sense tachyons go backward in time and seem to violate the sacrosanct principle of causality. As a response to this problem, early tachyon physicists suggested the so-called reinterpretation principle. This principle states (in one of its several versions) that emission of negative-energy tachyons can always be reinterpreted as absorption of positive-energy tachyons [6]. In this way the problem of tachyons endowed with negative energy disappears, or so it was claimed.

The belief in tachyons was to some extent motivated in the 'totalitarian principle' which can be considered a variant of the better known principle of plenitude with roots in Leibniz's philosophy [13]. Essentially, it is the claim that what is allowed by the laws of nature must actually exist.[3] In a paper of 1969, Olexa-Myron Bilaniuk and George Sudarshan explicitly related the search for tachyons, and also for magnetic monopoles, to what they called Gell-Mann's totalitarian principle [6]:

> There is an unwritten precept in modern physics, often facetiously referred to as Gell-Mann's totalitarian principle, which states that in physics "anything which is not prohibited is compulsory." Guided by this sort of argument we have made a number

---

[2] For the extensive tachyon literature, see e.g. [11] and [12]. The latter source lists more than 300 papers, Strum's not among them.

[3] See [14] for a critical analysis of the totalitarian principle and its historical sources. As pointed out in this paper, the common association of the principle with Murray Gell-Mann is unfounded. See also Robert C. Hovis, *Principles and the Development of Physical Theory: Case Studies*. Unpublished PhD dissertation, Cornell University, 1994. Hovis' dissertation refers to Strum's early ideas about superluminal signals.

of remarkable discoveries, from neutrinos to radio galaxies … If tachyons exist, they
ought to be found. If they do not exist, we ought to be able to say why not.

Tachyon physics makes up an interesting episode in the history of modern physics,
but it was a relatively short one, a parenthesis only. After about two decades, the
majority of physicists lost interest in the subject. Not only did the many searches for
tachyons lead to nothing, theoretical examinations also indicated that the paradoxes
inherent in the theory remained. Even more important, it seemed impossible to
construct a tachyon theory consistent with the requirements of quantum mechanics.
Latest by the turn of the century, the consensus view had returned to what it was
eighty years earlier: particles or signals travelling faster than light do not exist.

## 3. The velocity of light in vacuum

If only in retrospect, a new constant of nature entered physics in 1676 when the
Danish astronomer Ole Rømer, working in Paris, concluded from observations of the
moons of Jupiter that light propagates at a finite speed [15]. Half a century later, in
1728, James Bradley found from studies of stellar aberration that the speed is close to
$c = 300,000$ km/s and the same throughout the universe. Moreover, it was established
that the speed of light is the same for all colours or frequencies. At the time $c$ was
considered an interesting quantity, but not more interesting or fundamental than, for
example, the velocity of sound. Nor was it considered a limiting velocity, witness
that Pierre-Simon Laplace in the late eighteenth century argued that the force of
gravity propagates at a speed about 8 million times the speed of light ($v \sim 10^{15}$ m/s),
hardly distinguishable from instantaneity [16, p. 34].

The situation only changed in the second half of the nineteenth century when a
connection to laboratory physics was established with Maxwell's electromagnetic
theory of light and Heinrich Hertz's subsequent confirmation of it. According to
Maxwell's theory as propounded in his seminal 1865 article "A dynamical theory of
the electromagnetic field," the velocity of light was given by two constants
associated with the electromagnetic ether, today called the permittivity and
magnetic permeability of the vacuum: $c = 1/\sqrt{\varepsilon_0 \mu_0}$. Whereas the velocity was
traditionally an empirical constant, something which could be measured, since 1983
it has been *defined* conventionally as exactly $c = 299,792,458$ m/s. That is, the velocity
of light will forever have this value – or at least, until some other definition is agreed
upon. The length unit one metre is similarly defined as $1/c$, the distance that light
passes in one second.

By the late nineteenth century the speed of light was a central problem in the
so-called 'electromagnetic world view,' an ambitious attempt to reduce all of physics to the equations of electromagnetism instead of those of mechanics [17, pp. 105-119], [18, pp. 227-245]. Not only was the speed recognized to be a universal constant, it was also an invariant in the sense that it had the same value independent of the state of motion of the light source. This fundamental property, a postulate of Einstein's theory of relativity, was first stated by Henri Poincaré on the basis of his slightly earlier electron theory. Moreover, according to Poincaré the speed of light was a maximum speed for the propagation of any physical signal or particle. In an address of 1904 to the St. Louis Congress of Arts and Science he said about the new dynamics that it "will be characterized above all by the rule, that no velocity can exceed the velocity of light" [19, p. 293].

Poincaré's rule was broadly accepted by fin-de-siècle physicists although it did not follow unambiguously from the period's electron theories. These, it should be noted, referred to positive as well as negative unit charges. A minority of specialists in electromagnetic theory considered, more or less seriously, the possibility of superluminal particles [20]. The first was perhaps the British amateur physicist Oliver Heaviside, who in a series of papers starting in 1889 examined theoretically a moving charge and its associated increase in electromagnetic mass [21], [22, pp. 124-126], [23]. He saw no reason to exclude from his analysis a charge moving in vacuum with a velocity greater than the speed of light, $v > c$.

There are several references to superluminal velocities in Heaviside's *Electromagnetic Theory*, where he discusses, for example, "an electron [which] is jerked away from an atom so strongly that its speed exceeds that of light" [21, pp. 164-165, originally in *Nature* 1903]. Elsewhere [21, p. 69] he referred to some high-voltage experiments made by the American physicist John Trowbridge, on which he commented:

> It seems very probable that in his experiments electrons do have speeds given to them exceeding that of light … An electron moving much faster than light does will draw after it other slowly-moving electrons of the same sign, instead of repelling them.

According to Heaviside, superluminal motion of this kind was mathematically as well as physically allowed. Other British physicists disagreed, arguing that if accelerated to the velocity of light the energy of the charge would become infinite, and "so it would seem to be impossible to make a charged body move at a greater speed than that of light" [24], [22, p. 124].

The German physicist Arnold Sommerfeld was another physicist who seriously considered the possibility of faster-than-light electrical particles, which he did in more detail than Heaviside [1, pp. 163-166]. Of course, Sommerfeld is today much better known for his later and very important contributions to quantum theory, theoretical spectroscopy and atomic structure. While a professor at the Technical University in Aachen he was deeply engaged with the mathematical aspects of the electromagnetic electron which he examined in papers of 1904-1905 [25], [26]. There were at the time basically two rival models of the electron. Max Abraham in Germany pictured the electron as a rigid sphere with charge uniformly distributed on either the surface or over the entire volume. The alternative favoured by Hendrik A. Lorentz in the Netherlands was not rigid but deformable as the electron would contract in its direction of motion at high speed, namely as $l = l_0\sqrt{1 - v^2/c^2}$, what came to be known as the Lorentz contraction.

Sommerfeld's calculations ruled out superluminal motion of Lorentz's electron and also of Abraham's rigid electron with surface charge. On the other hand, if the charge was distributed over the volume the Abraham electron might move faster than light in vacuum although in a way that was highly unusual from a physical point of view. Not only would it require force to slow down a superluminal electron, an electron in this state of motion also seemed to have a negative mass! For a brief while Sommerfeld thought that experiments made by Friedrich Paschen on gamma rays provided evidence in favour of superluminal electrons. As he stated in a letter to Lorentz of 14 December 1904, "From what Paschen has told me about his most recent experiments I must assume that … motion faster than light really exist … [but] I am not quite sure" [27, p. 152]. About two months later Paschen wrote to Sommerfeld: "I would be reassured if the theory indicates that by exceeding the speed of light the [magnetic] deflection must be reduced. Should that not be the case it would disprove that gamma rays are of the same nature as cathode rays" [27, pp. 239-240].

Sommerfeld's fascination of faster-than-light particles did not last long. In a paper of 1905 written shortly before Einstein's on relativity theory, he expressed serious doubt about the physical reality of the mathematically allowed superluminal electrons. He now concluded that "motion faster than light is in no way a physically real phenomenon" [27, p. 153]. The leading Göttingen physicist Emil Wiechert, an expert in electron theory (but today best known as a pioneering seismologist), was less sure. In part inspired by Sommerfeld's work he analysed in early 1905 how electrons should be structured to move faster than light, something he considered a realistic possibility [28], [29].

## 4. Einstein's theory and Tolman's paradox

Contrary to Heaviside, Sommerfeld, Poincaré, Wiechert, Lorentz and other electron theorists, Einstein did not consider material particles to be made up of electromagnetic fields. His new theory of relativity dealt with any kind of matter whether electrical or not. Although he did write of electrons in his epoch-making 1905 paper in *Annalen der Physik*, he made it clear that his 'electrons' had nothing in common with what other physicists associated with the term [30]. For him it was just a "material point" or "a particle endowed with an electric charge," as he stated in a paper two years later. Aware of the German electron theorists' discussions of hypothetical faster-than-light particles, in his later article Einstein referred twice to the possibility but only to reject it as "meaningless" [31], [32, pp. 56-61]. As he showed, the kinetic energy of an electron moving at speed $v$ would become infinite for $v = c$. The mass of a body increases with its velocity according to $m = m_0\gamma$, where $\gamma$ is the Lorentz factor

$$\gamma = \frac{1}{\sqrt{1 - v^2/c^2}}$$

Consequently, for $v > c$ the mass becomes imaginary and therefore unphysical. In another part of the paper Einstein proved that if two velocities $v_1$ and $v_2$ are added, the resulting velocity is given by the formula

$$u = \frac{v_1 + v_2}{1 + v_1 v_2/c^2}$$

Thus it is impossible to pass the $v = c$ barrier by adding subluminal or even luminal velocities (for $v_1 = c$ and $v_2 = c$, the result is $u = c$). "Velocities greater than that of light have … no possibility of existence," Einstein summarily stated at the end of his article [30, p. 920], [5, pp. 236-238].

In his lesser known 1907 paper in *Annalen der Physik*, Einstein introduced a new and powerful conceptual argument against physical signals propagating with a superluminal velocity, namely that such motion would violate the principle of causality. This he illustrated by means of a thought experiment from which he concluded [31, p. 381]:

> We would have to consider as possible a transfer mechanism whose use would produce an effect which *precedes* the cause (accompanied by an act of will, for example). Even though, in my opinion, this result does not contain a contradiction from a purely logical point of view, it conflicts so absolutely with the character of all our experience, that the impossibility of the assumption [$v > c$] is sufficiently proved by this result.

Einstein repeated almost verbatim the causality objection in another 1907 paper published in Johannes Stark's *Jahrbuch der Radioaktivität und Elektronik* [33]. In a correspondence with Wilhelm Wien, he once again emphasized his disbelief in signals or particles transmitted with a speed greater than that of light. Wien, a distinguished professor at the University of Würzburg, was a leading electron theorist who in 1911 became a Nobel laureate for his work on heat radiation. In one of his letters to Wien, Einstein wrote: "The impossibility of superluminal signals follows with certainty" [34, p. 71]. For a popular audience he returned to the impossibility in *The Evolution of Physics* co-authored by Leopold Infeld and first published 1938. Imagine, he said, that we move at a speed exceeding that of light. Then we would experience events from the past in the reversed temporal order to which they really occurred: "The train of happenings on our earth would appear like a film shown backward, beginning with a happy ending" [35, p. 177].

Einstein's causality argument against faster-than-light signals was noted early on in the growing anti-relativity literature. Citing from the 1907 paper, one of the anti-relativists, an Austrian engineer by the name Leo Gilbert, found it preposterous that Einstein had no *logical* objections to cause-effect reversal. To make his point, he presented a more common-sense example [36]: "The farmer's wife finds an egg in the stable. In a week she will go to the market and by the hen that is to lay the egg she has already found. Who can object to that? Einstein agrees with it."[4] Expectedly, also philosophers took notice of the problem. Thus, at a symposium on time in the Aristotelian Society the British philosopher Herbert Wildon Carr discussed the claim that "no movement of translation can exceed the velocity of light, because if it could there would be observers for whom the effect would happen before the cause" [39, p. 129].

While Gilbert deliberately misrepresented Einstein's argument as evidence that relativity theory was nonsensical, the American chemist and physicist (and later pioneering cosmologist) Richard Tolman got it right. In a monograph of 1917 he described what came to be known as 'Tolman's paradox,' namely that the assumption of superluminal signals implies an effect preceding its cause. Tolman's version of the paradox was almost identical to Einstein's, if better known and a bit more elaborate. Referring to the cause-effect reversal, he wrote: "Such a condition of affairs might not be a logical impossibility; nevertheless its extraordinary nature might incline us to believe that no causal impulse can travel with a velocity greater than that of light" [40, p. 55]. As Tolman emphasized, the impossibility was restricted to communication of some sort where a signal carrying information causally connects one event with another. He restated the argument in the important and widely read textbook *Relativity, Thermodynamics and Cosmology* [41, p. 27]:

> Retaining our ideas as to cause and effect as being essentially valid for macroscopic considerations, it can be shown that causal impulses cannot be transmitted with a velocity greater than light, since it would then be possible to find systems of coordinates in which the effect would precede the cause.

---

[4] Based on Gilbert's booklet [37] reviewed in *Nature* **93** (1914): 56-57. Gilbert did not mention the superluminal signals that Einstein was concerned with. He suggested, quite wrongly, that the theory of relativity justified backwards causation and should therefore be dismissed. On Gilbert and early anti-relativism, see [38, pp. 84-85].

Tolman's paradox was much discussed in later tachyon physics, now sometimes under the more colourful but somewhat unfortunate name 'tachyonic antitelephone' [42]. Using a hypothetical telephone system based on superluminal signals it would be possible to deliver a message to the past. The message would be heard before it was sent.

The backwards causation implied by superluminal signals was commonly known in the early 1920s when laypersons were exposed to the many strange consequences of Einstein's theory of relativity. Readers of *Punch*, a satirical British weekly magazine, could in 1923 enjoy a limerick that offered a new version of Tolman's paradox:

> There was a young lady named Bright
> Whose speed was far faster than light.
> She went out one day
> In a relative way
> And returned on the previous night.

The limerick, soon to become famous, appeared anonymously but was composed by Arthur Reginald Buller, a distinguished British-Canadian botanist.[5]

With the acceptance of the special theory of relativity almost all physicists agreed that superluminal particles or signals cannot possibly be part of nature's fabric. As emphasized by Hans Thirring, a distinguished physicist at the University of Vienna and an expert in Einstein's theory, "actions cannot propagate with velocity greater than that of light, and it follows naturally that neither can material bodies travel at such velocity" [45, p. 77]. The consensus view was repeated in later textbooks which often cited the cause-effect objection as convincing evidence. To mention but one example, a version of Tolman's paradox appeared in the Danish physicist Christian Møller's widely read textbook on relativity theory from 1952 with the comment: "Obviously … in nature no signals can exist which move with a velocity greater than the velocity of light relative to any system of inertia. This represents a general statement regarding the fundamental laws of nature" [46, p. 53].

---

[5] *Punch* **165** (19 December 1923), 591. On the origin and history of the verse, see [43] and [44, p. 11]. In the original form of the limerick, the last lines were "She set out one day / In a relative way / And returned home the previous night."

## 5. An unknown Russian physicist

Lev Yakovlevich Strum, 1890-1936. If the name doesn't ring a bell it is pardonable, for the Russian–Ukrainian Strum (also spelled Shtrum) is largely invisible in the scholarly literature on the history of Russian physics in the interwar period. None of the major English-language works in this well-researched area of history of science as much as mentions him, e.g. [47], [48], [49], [50], [51], [52].

Born into a middle-class Jewish family in an Ukrainian town in 1890, Strum first studied mathematics in St. Petersburg (Petrograd) from where he graduated in 1914 [3], [53]. After the Communist Revolution and its aftermath, the Civil War, he obtained a position at the Kyiv Polytechnic Institute's department of physics, whose first director, the Polish-born later academician Alexander G. Goldman (also a Jew), supported Strum's career. In 1932, after reorganization and now with Strum as its head, the department was turned into a branch under the Ukrainian Academy of Sciences. Four years later it was renamed the Institute of Physics of the Ukrainian Academy of Sciences.[6] In 1927 Strum defended a doctoral dissertation on the quantum theory of X-rays. Another subject he investigated, and which will be dealt with below, was the special theory of relativity. In 1932 he was appointed professor of theoretical physics at Kyiv State University.

Strum was at the time a respected if somewhat peripheral figure in Russian physics and part of the network that included much better-known physicists such as Yakov Frenkel, Lev Landau, George Gamow, Dmitrii Ivanenko and Matvei Bronstein (Bronshtein). However, the centres of theoretical physics in the Soviet Union were unquestionably Leningrad, Kharkiv and Moscow. Compared to these centres the Kyiv institute was more provincial and enjoyed less international reputation. Before entering the University of Leningrad (formerly St. Petersburg) in 1926, young Bronstein studied in Kyiv where he worked on some of the same research topics that Strum was engaged with [49, pp. 16-18].

---

[6] For a brief history of the Kyiv Institute of Physics, see http://www.iop.kiev.ua/en/history/ . Accessed on 20 March 2024.

Strum attended in May 1929 the First All-Union Conference on Theoretical Physics held in Kharkiv with other participants including Ivanenko, Frenkel, Gamow, Landau and the German quantum physicists Walter Heitler and Pascual Jordan (see [54] with a photograph of Strum, Jordan and other physicists). He also participated in the Third All-Union Conference held in Kharkiv 1934 which had Niels Bohr as its big attraction and where he met with Bronstein and many other physicists. This conference was chiefly organized by Landau, who since 1932 had been head of the Ukrainian Physico-Technical Institute's strong school in theoretical physics [52, pp. 92-98], [55]. According to one source (the reliability of which is doubtful, however), "During one of the conferences abroad, Shtrum met Albert Einstein and received an autographed photograph from him" [56].[7]

A productive and versatile researcher, Strum published scientific articles not only in Russian but also in internationally recognized journals such as *Zeitschrift für Physik*, *Zeitschrift für physikalische Chemie*, *Physikalische Zeitschrift* and *Nature*. These papers were abstracted in the physics volumes of *Science Abstracts*, a leading abstract journal issued by the Institute of Electrical Engineers and the Physical Society of London. Strum's publications in Russian appeared in part in the *Uspekhi Fizicheskikh Nauk* (Advances in Physical Sciences) established in 1918 and in part in an in-house journal published by the Kyiv physics institute between 1929 and 1936 [53]. One of his contributions to *Uspkekhi* was a Russian translation of Erwin Schrödinger's comprehensive article on wave mechanics published in *Physical Review* [58].

Among Strum's works on relativity theory in the mid-1920s was a paper on the controversial ether-drift experiments by the American physicist Dayton Clarence Miller, who had announced a small but positive result disagreeing with the theory of relativity [59]. Instead of questioning Miller's data, as Einstein did,[8] Strum proposed a non-relativistic hypothesis according to which the velocity of light $c$ depends on the velocity of the light source $v$ as $c = c_0 + \alpha v$, where $c_0$ denotes the speed of light for a source at rest [62]. He stated that the coefficient $\alpha$ had to lie in the interval $0 < \alpha < 1$ "since $c$ cannot be less than $c_0$ and not greater than $c_0 + v$." With $\alpha = 1/3$ he believed to have offered a credible explanation of Miller's observational data.

---

[7] As far as I know, Strum never met Einstein and never participated in conferences outside the Soviet Union whether in Leiden or elsewhere. The claim seems to have its origin in Strum's daughter Elena Lvovna, who at the age of 95 recalled "how her father proudly showed her a photograph with the autograph of his idol A. Einstein, whom he met at a conference in Leiden" [57].

[8] It was on this occasion that Einstein commented, "Subtle is the Lord, but malicious He is not" (Raffiniert ist der Herr Gott, aber boshaft ist er nicht). For this quote and Einstein's response to Miller's series of experiments, see [60] and [61, pp. 287-290].

Strangely, Strum did not comment on the obvious disagreement of his hypothesis with Einstein's theory of relativity.

Shortly before his work on the velocity of light, Strum published, also in *Zeitschrift für Physik*, a detailed paper on the intensity of spectral lines on the basis of the Bohr-Sommerfeld quantum atomic theory [63]. He obviously mastered the technical details of this theory and had carefully studied the works of Bohr, Sommerfeld, Planck and other leaders of the old quantum theory. Among other things, he explained quantitatively what Bohr had suggested qualitatively in his 1913 masterpiece, namely that there are many more hydrogen lines in the Balmer series from stellar spectra than those obtained in the laboratory [64, p. 62].

Strum suggested in 1928 a generalization of Planck's fundamental law of blackbody radiation which implied a slight revision of the coefficient in the Stefan-Boltzmann law that agreed better with recent measurements [65]. When this paper is worth noticing it is for two reasons. First, it shows that at the time Strum was in personal contact with the German physicist Rudolf Ladenburg, one of the leading figures of quantum physics. Second, the paper had some impact as it was cited by, among others, the American physicist Raymond Birge in an important review of the constants of nature [66]. However, Birge disagreed with Strum's suggestion that the Planck formula was in need of revision. Eighteen-year old Subrahmanyan Chandrasekhar, still in India but on his way to become a famous astrophysicist and eventually a Nobel laureate, was directly inspired by Strum's paper. In 1929 he proposed with youthful confidence an ambitious new theory of quantum statistics for gases and photons that he saw as a further generalization of Strum's theory [67].[9] In fact, the very first word in his paper was "L. Strum."

In other papers from the period Strum investigated the atomic nucleus which until the early 1930s was universally believed to consist of protons and electrons [68], [69]. In a contribution of 1928 he calculated on this basis the size of atomic nuclei, stating that the diameter of the helium nucleus (four protons and two electrons) was $1.4 \times 10^{-14}$ cm. This paper was cited by other experts in early nuclear physics such as Ernest Rutherford in England, Fritz Houtermans in Germany, Fritz Kohlrausch in Austria, and James Bartlett in the United States. After the discovery of the neutron and the application of quantum mechanics to the nuclear domain, Strum suggested a mechanism for possible transformations between protons and neutrons by means of photons in relation to a theory suggested by his compatriot, the

---

[9]  Curiously, this paper is missing in Chandrasekhar's list of publications as presented in *Journal of Astrophysics and Astronomy* **17** (1996): 269-298. It was his second paper, the first one being a paper published in 1928 in *Indian Journal of Physics*.

Ukrainian-born Dmitrii Ivanenko [70], [71]. However, whereas Ivanenko suggested hypothetical massive exchange particles, Strum considered only photons as in the reactions $n + h\nu \rightarrow p^+ + e^-$ and $p^+ + h\nu \rightarrow n + e^+$. Moreover, the mass value that he adopted for the neutron, 1.0067 on the basis of $O^{16} = 1$, implied that it was a proton-electron compound and not an elementary particle [72, pp. 224-227].

## 6. Death and literary resurrection

The *Nature* paper on nuclear physic was one of Strum's last works, for only two years later he tragically ended his life before a firing squad together with 36 other Ukrainian scientists and intellectuals. Like other victims of Stalin's Great Terror, he was forced under torture to plead guilty in the charges of being an 'enemy of the people' and an agent of the Trotskyist underground network. He was arrested by agents from the secret police NKVD on 3 March 1936 and executed the same year on 22 October in the Lukyanivska Prison in Kyiv [3], [53], [73, p. 202]. On the top of that, Strum's wife was exiled to the far-away Arkhangelsk region and later arrested. She spent eight years in a labour camp and when she was released in 1953, she was forbidden to live in large cities. Strum's fate is not generally known such as indicated by a quotation from a book written the historian Alexander Vucinich [50, p. 223]:

> The Stalinist terror in 1936-38 took a heavy toll on … leaders of Einsteinian studies. Seven top experts in the theory of relativity – two philosophers (S. Semkovskii and B. M. Hessen) and five physicists (M. Bronshtein, V.K. Frederiks, V.A. Fock, Iu. B. Rumer, and L.D. Landau) – were sent to Stalinist prisons, from which four did not return.[10]

Notice that Strum is conspicuously absent from Vucinich's list.

In 1938 the younger Bronstein, a brilliant theoretical physicist, suffered the same cruel fate as Strum, such as did many other Soviet physicists and astronomers [49], [17, pp. 240-244]. Among them was the low-temperature physicist Lev Shubnikov, who worked at the Physico-Technical Institute in Kharkiv and was arrested in August 1937 on largely the same charges as Strum and Bronstein. He was executed a few months later [74]. Landau too was in great danger, but he avoided a death sentence after having spent a year in jail. After Strum's arrest, his chair in theoretical physics at the Institute of Physics in Kyiv was in 1936-1938 occupied by the American-Jewish physicist Nathan Rosen, a collaborator of Einstein and famous for his co-authorship of the 1935 EPR (Einstein-Podolsky-Rosen) thought experiment in quantum mechanics. A.G. Goldman, Strum's friend and mentor, was yet another Jewish victim of the Stalinist purge. Accused of supporting terrorist and anti-state activities he was arrested in 1938 and deported for five years to Kazakhstan.

---

[10]  Sergei Semkovskii (1883-1937) was a prominent Ukrainian philosopher who supported relativity theory as consistent with Marxist dialectical materialism [47, 119-122]. He was arrested on 3 March 1936. Strum was acquainted with Semkovskii with whom he discussed questions of natural philosophy [3]. B.M. Hessen refers to the Russian-Jewish physicist Boris M. Hessen (or Gessen) who is primarily known for his Marxist interpretation of the history of science [47, 185-188]. He was executed on 20 December 1936.

Not only was Strum executed on false charges, the Soviet authorities also turned him from a person into a 'non-person' by erasing the memory of him and destroying what they could of his scientific and other works. It was as if Lev Strum had never existed. Given his scientific papers published in Western journals one might assume that he was kept alive, so to speak, outside the Soviet Union, but this was not the case. He was as unknown there as he was in Stalin's Russia. When he eventually was resurrected it was, strangely, as a literary figure not immediately recognized as the real Lev Strum.

After numerous difficulties and problems with the Soviet censorship, the Russian writer Vasily S. Grossman (1905-1964), like Strum a Ukrainian Jew, published in 1960 the novel *Life and Fate* which is today considered a masterpiece in twentieth-century literature. It has even been compared to Leo Tolstoy's classic *War and Peace*. The novel's key figure is a certain Viktor Pavlovich Strum, a brilliant physicist whose life and science during the difficult 1930s and 1940s are described in fascinating details. Apparently a fictional character, Grossman introduced Viktor Strum as a literary substitute of the deceased Lev Strum already in his previous novel *Stalingrad* of which *Life and Fate* is a kind of sequel. While a young man Grossman knew the real Strum, who may have been one of his teachers in Kyiv and to whom he was deeply indebted [75, p. xxii].[11]

Trained as a chemical engineer, Grossman maintained an interest in science and in physics in particular. Like Lev Strum, he was an admirer of Einstein. This is evidenced in the many passages in both *Stalingrad* and *Life and Fate* referring to physicists and areas of physics and chemistry. Thus, in the early part of the latter novel, Viktor says, out of the blue, "Lyuda, you remember Prout's hypothesis?" He then goes on to explain what this hypothesis is about [76, pp. 65-66]:

---

[11] This is according to Robert Chandler, the translator of Grossman's novels. See also "Literature and reality," https://jordanrussiacenter.org/blog/literature-and-reality-with-robert-chandler-event-recap. Accessed on 20 March 2025. In its original and censored version published in 1952 *Stalingrad* carried the title *For a Just Cause*. See details in [73] where Viktor Strum is described as Grossman's alter ego. For the connection to Lev Strum, see also [77]. Until recently historians of literature believed that Landau, and not Strum, was the model of Grossman's fictional character.

What happened with Prout is that he arrived at a correct hypothesis largely because of the gross errors that were current in the determination of the atomic weights. If the atomic weights had already been determined with the accuracy later achieved by [Jean Baptiste] Dumas and [Jean Servais] Stas, he'd never have dared hypothesize that they were multiples of hydrogen. What led him to the correct answer was his mistakes.[12]

There is also a description of an unspecified experiment conducted in Cambridge "shortly before the war" that confirmed the theory of relativity "in certain extreme conditions" [76, p. 261]:

The success of this experiment was the theory's most brilliant triumph. To Viktor, it seemed as exalted and poetic as the experiment on relativity which confirmed the predicted deviation of a ray of light from a star passing through the sun's gravitational field. Any attack on this theory was quite unthinkable – it would be like a soldier trying to rip the gold braid off a field-marshal's shoulders.

In yet another passage, Viktor Strum reflects on how his own work will be received by his colleagues in physics [76, pp. 344-345]:

Previously, it hadn't even occurred to him to share his thoughts with anyone else. He wanted to see Sokolov and write to Chepyzhin; he wondered what Mandelstam, Joffe, Landau, Tamm, and Kurchatov would think of his new equations; he tried to guess what response they would evoke in his colleagues both here in the laboratory and in Leningrad. He tried to think of a title for his work. He wondered what Bohr and Fermi would think of it. Maybe Einstein himself would read it and write him a brief note.[13]

Apparently unaware that Viktor Strum was modelled on Lev Strum, the Soviet editors nonetheless objected to the name of the character because it was Jewish. Although pressured to find another name, or to downplay the role of Strum considerably in his novel, Grossman insisted on keeping the name if wisely providing it with fictional first names [77].

---

[12] According to the physician and chemist William Prout's famous hypothesis of 1815, the atomic weight of any element was a multiple of hydrogen's weight, that is, $A_x = nA_H$ with $n = 1, 2, \ldots$ . The relation was often interpreted as evidence that all chemical elements consists of hydrogen atoms. For details and historical context, see [78].

[13] Of the distinguished Soviet physicists referred to, Chepyzhin was modelled on Pyotr Kapitsa (1894-1984), who in *Stalingrad* is described as Viktor's inspiring teacher [75, pp. 170-174]. The others, appearing with their own names, were Sergey L. Mandelstam (1910-1990), Abraham F. Joffe (1880-1960), Lev D. Landau (1908-1968), Igor E. Tamm (1895-1971), and Igor V. Kurchatov (1903-1960). Kapitsa, Tamm and Landau received the Nobel Prize in physics. Pjotr L. Sokolov, an invented figure, appears in the novel as a mathematician in Viktor Strum's laboratory.

Although Strum was posthumously rehabilitated for his alleged crimes in 1956, three years after Stalin's death and now with Khrushchev in power, for a long time he remained a non-person rather than a person. It took until about 2010 before a few physicists, historians and literary critics called attention to his existence and role in Ukrainian (rather than Russian) history of science. In February 2018, a memorial plaque was installed on the house where Lev Strum and his family once lived in Kyiv, namely on 12 Malopidvalnaya Street in the central part of the city [57].

## 7. Strum on faster-than-light signals

At a conference in Kyiv in September 1921, Strum presented a report on the possibility of superluminal signals which he transformed into an article that appeared in *Zeitschrift für Physik* two years later [79], [80], [3]. It was followed by a couple of related papers in Russian and German. He introduced the article by suggesting that despite the fundamental nature of Einstein's general theory of relativity, the special theory should not be forgotten. With regard to the latter theory he noted, correctly, that "most writers consider the velocity of light in vacuum to be not only a universal constant but also the highest imaginable velocity." Strum begged to disagree.

After a critical review of the standard objections against superluminal velocities, Strum [79] concluded that they were not compelling. Signals moving at a speed greater than that of light did not necessarily contradict either the relativity principle or the law of causality. He considered two one-dimensional systems of inertia in which S′ moved relative to the system S at a subluminal velocity $v < c$. In the latter system he imagined a superluminal signal propagating from its origin in both directions. From this he found that "the time coordinate is negative only in system S′, while in the other system S, it attains a positive value." Moreover:

> For processes propagating with a speed faster than light in any way whatever, system S allows such velocity of rectilinear and uniform motion of another system S′ relative to S at which the time in system S′ runs for such processes in a direction opposite to the flow of time in system S.

As Strum cautiously pointed out in a footnote, "Of course, this is not about the real existence of such [superluminal] velocities but only shows that they are theoretically possible."

By means of a space-time diagram of the type introduced by Hermann Minkowski in 1908, Strum illustrated that superluminal signals could propagate outside the $v = c$ light cone which was normally considered unphysical. "The points outside the cone" he said, "correspond to velocities greater than that of light. For such velocities the direction of time can under certain conditions be seen as opposite to the [ordinary] direction of time." He summarized his analysis in two statements. First, that the hypothesis of superluminal velocities does not contradict the special theory of relativity. And second, that the concept of direction of time is itself relative in the sense that what in one system is before a later event may in another system be after this event. As Strum saw it, this dismantled the causality objection without denying the fundamental law of causality.

In a follow-up paper [81] he argued that whether outside or inside the light cone, the effect always comes after the cause, "but what is conceived as cause and effect can be different in different systems." According to Strum, there would be no causal paradox because a superluminal signal sent back in time can be interpreted as a signal moving forward in time. This has been seen as a version of the 'reinterpretation principle' of tachyon physics mentioned in Section 1 [3], [80]. In yet another paper on superluminal motion Strum discussed whether a wave with phase velocity greater than $c$ could carry information and thereby function as an 'antitelephone' signal. Concluding that this was not the case, he added: "Naturally, from this it does not follow that there cannot exist, apart from the phase wave, other velocities greater than that of light" [82].

Strum seems to have conceived the hypothetical faster-than-light signals to belong to a 'world' with no physical connection to the ordinary world characterized by velocities smaller than or equal to the velocity of light. The theoretically allowed signals travelling faster than light would always move in this way without ever crossing the light barrier $v = c$. In this respect his theory anticipated the much later theory of tachyons. On the other hand, Strum did not consider the possibility of massive superluminal particles and therefore ignored the dynamical problem of how to interpret the imaginary mass of such particles. His work of 1923 was innovative, but the claim that Strum "suggested the existence of tachyons" lacks justification [80]. Interested only in the theoretical possibility of signals moving faster than light he also did not comment on how to detect such signals. Philosophically speaking, he seems to have been closer to instrumentalism than realism.

By the mid-1920s special relativity theory was fully accepted and no longer an active research area. Interest had moved to Einstein's general theory and, to an even greater extent, to quantum and atomic physics. Very few physicists thought that the question of superluminal velocities in special relativity was worth looking at. Perhaps for this reason, Strum's papers failed to attract attention such as is reflected in their poor citation records.[14] Strum may have hoped that Einstein read his paper in *Zeitschrift für Physik*, such that he did with the Russian physicist Alexander Friedmann's important 1922 paper on relativistic cosmology published in the same journal [83] [84, pp. 169-185]. However, in all likelihood he did not. There is no trace of Strum in *The Collected Papers of Albert Einstein* or in any other reliable source on Einstein.

Robert Bass, a young Austrian physicist, covered some of the same ground as Strum in papers from 1926 and 1927, but without citing – and possibly without knowing about – the work of his Ukrainian colleague. Much like Strum, Bass argued that there is no contradiction between special relativity and superluminal signals so long that the latter are not carried by massive particles. Referring to Einstein's 1907 paper, he stated [85]:

> Apart from the paradoxes, which mostly belong to epistemology rather than physics, there are no physical reasons to assume that the velocity of light is an upper limit for the propagation of actions … The possibility of superluminal actions has so far not been strictly refuted.

Contrary to Strum, Bass was willing to dispense with the law of causality, which he regarded as an empirical hypothesis and not a logical necessity. Despite his defence of superluminal motion of physical signals, Bass believed that velocities greater than that of light posed a problem for free will. This he illustrated with a thought experiment, a variant of Tolman's paradox, which he owed to his professor Hans Thirring:

> An observer on the Earth sends a signal faster than that of light by, for example, pressing a knob. On planet Neptune an observer receives the signal and answers it at once; the man on the Earth receives the reply before he has even sent the question. This leads to a conflict with the freedom of will.

Bass' argument that superluminal communication conflicts with free will was contradicted by Josef Würschmidt, a German-Argentine physicist who argued at length against the conclusion of Bass and Thirring [86]. While Strum [81] responded to Bass's paper, neither did Bass nor Würschmidt refer to Strum's earlier analysis of motion with a velocity greater than that of light. His work in this field was effectively forgotten years before he physically vanished from the scene in 1936. In a recent paper two physicists point out that "his [Strum's] works were removed from Soviet libraries and destroyed," and then continue: "As a result, the tachyon hypothesis fell into oblivion for many years" [87]. However, there is no connection between the two events. After all, Strum's papers on relativity theory were freely available in German journals. This could not be changed, not even by the diligent Soviet censors. Strum's hypothesis fell into oblivion right after it was proposed for the simple reason that physicists, whether in the East or the West, found it to be of no scientific interest. The same was the case with Bass' similar hypothesis.

---

[14] Google Scholar reports no citations at all in the period 1923-1940 to Strum's 1923 paper. Web of Science reports a single citation, but this is a self-citation of 1930. The paper was possibly cited in Russian journals not covered by these data bases but even so it remained largely invisible until about 2010.

## 8. Conclusion

This paper describes two separate but still related topics in the history of twentieth-century physics, namely the early development of faster-than-light theories and the contributions on this subject by the Kyiv physicist Lev Strum. While aspects of the first area have previously been covered in the historical literature, this is not the case with the second area. Indeed, when Strum turns up in modern literature it is predominantly as the semi-fictional figure Viktor Strum in Grossman's novel *Life and Fate* and not as the real Ukrainian-Jewish physicist Lev Yakovlevich Strum.[15] As Grossman's work throws new light on an episode in the development of science, and for that reason is relevant to historians of science, so is it the case with some other fiction literature whether novels, drama or poems. As I have argued in an earlier paper, another example of this kind relating to the history of twentieth-century physics is Bertold Brecht's famous play *The Life of Galileo* [88].

Apart from its focus on Strum the present article calls attention to the few earlier pre-relativistic suggestions that electrons might in some cases travel faster than light in vacuum. Although no genuinely superluminal signal or tachyonic particle has ever been detected, physicists continue to discuss the possibility, but now mostly within the very different context of quantum mechanics [89]. However, these modern discussions are clearly outside the chronological and conceptual framework of this paper. The same is the case with the tachyons that turned up in the early phase of string theory at about the same time that superluminal particles were discussed by tachyon physicists. The 'tachyons' that string theorists wanted to eliminate from their theory were not particles moving faster than light but particle states with an imaginary rest mass [2, pp. 292-305], [90].

---

[15] I have recently become aware of arguments based on textual evidence which question the hypothesis that Grossman modelled the fictional Viktor Strum on the real-life Lev Strum [91]. The hypothesis may be right, but there are no compelling reasons to elevate it to a fact.